%% file: main.tex
\documentclass{article}
\usepackage[utf8]{inputenc}
\usepackage{authblk}
\input{packages.tex}

\title{Performance Evaluation of Dynamic Scaling on MPI}
\author[1]{Masatoshi Hanai}
\author[1]{Georgios Theodoropoulos}
\affil[1]{\small{Southern University of Science and Technology, Shenzhen, China}}
\date{}

\begin{document}
\maketitle
\begin{abstract}
Dynamic scaling aims to elastically change the number of processes during runtime to tune the performance of the distributed applications.
This report briefly presents a performance evaluation of MPI process provisioning / de-provisioning for dynamic scaling by using 16 to 128 cores.
Our dynamic scaling implementation allows the new MPI processes from new hosts to communicate with the original ones immediately.
Moreover, it forbids the removing MPI processes to communicate with others as well as gets the information whether the host node can be terminated or not.
Such a simple feature is not supported as a single-line API in MPI-2 such as \texttt{MPI\_Comm\_spawn()}.
We provide our implementation as a simple library\footnote{It is publicly available in https://github.com/masatoshihanai/MPIDynamicScaling} to extend a non-dynamic-scalable MPI program into a dynamic-scalable one by adding only several lines of codes. 
\end{abstract}

\section{Introduction}
\emph{Dynamic scaling}
aims to elastically change the number of processes at runtime to tune the performance of the distributed applications on demand.
It has become increasingly crucial for distributed computation to utilize elastic computational resources such as the cloud.
The cloud allows the flexible control of virtual machines (VMs), depending on various requirements, e.g., the time and financial constraints.
\emph{Unreliable VM instances}, such as Spot Instances in AWS~\cite{spotinstance} and Preemptible VMs in GCE~\cite{preemptible}, provides more economical computational power but requires the dynamic scaling for handling the forced termination of VMs.

In MPI-2~\cite{mpi2}, the dynamic scaling is achieved by invoking new child processes from a certain process of the running MPI program, e.g., \texttt{MPI\_Comm\_spawn()}.
However, MPI-2 only provides primitive APIs to create or delete processes, and thus, these do not satisfy the following features:
\begin{itemize}
\item Communicate among the original MPI processes and the additional ones.
\item Forbid an invalid communication to the deleted MPI processes.
\item Host management to determine whether the hosts of the removal MPI processes can be terminated or not.
\end{itemize}
In other words, we require to establish a new intra-communicator after creating or removing MPI processes and to manage the host information.

In this report, we present an implementation for dynamic scaling and evaluate its performance.
We provide the implementation as a simple library, which enables us to easily extend the non-dynamic-scalable MPI program into the dynamic-scalable one by adding a few lines of codes.
\section{Background}
We review the MPI process management and communicator mechanisms.

\medskip\noindent
\textbf{\textit{MPI Communicator}}: There are two types of communicators in MPI: intra-communicator and iter-communicator.
Intra-communicator provides the basic communication among the MPI processes within a single MPI communicator, e.g., \texttt{MPI\_COMM\_WORLD}.
The processes within the same communicator are managed by identical rank (e.g. it can be obtained via \texttt{MPI\_Comm\_rank()}).
In MPI, multiple intra-communicators are allowed to separate all the MPI processes into several groups (e.g., the intra-communicator can be split via \texttt{MPI\_Comm\_split()}).
Iter-communicator provides communication between such separated groups.

\medskip\noindent
\textbf{\textit{Dynamic MPI Process Creation}}: In MPI-2, new processes are dynamically created by invoking an MPI program as children of a certain process of the running MPI processes (e.g., \texttt{MPI\_Comm\_spawn()}). 
The created processes are involved in the independent intra-communicator from the original ones.
Thus, the communication among the new and original processes is not allowed immediately.

\medskip\noindent
\textbf{\textit{MPI Communicator Integration and Separation}}: The intra- and inter-communicators can be integrated or separated by the following MPI functions:
\texttt{MPI\_Intercomm\_merge()} merges multiple groups in an inter-communicator and generates a new single intra-communicator among the processes of the groups; 
\texttt{MPI\_Comm\_split()} splits a single intra-communicator into several ones.

\section{Dynamic Scaling Implementation on MPI}
In this section, we present our dynamic scaling implementation, which does not only generate/remove MPI processes but also establish the intra-communication for all the processes after scaling. 
We first provide an implementation to scale out, namely, the new processes are added during runtime.
Then, we provide one to scale in, namely, the processes are removed from the running processes.
\newpage
\subsection{Scale Out Implementation}
Code~\ref{lst:scaleout} shows the implementation of scaling out in the original processes.
\begin{lstlisting}[caption={Scale Out Implementation in C++ (Original Processes)},label={lst:scaleout}]
void scaleOut(MPI_Comm oldComm, int numAdd, 
    string childProgram, MPI_Comm& newComm, 
    vector<string> hosts) {
  MPI_Info info = MPI_INFO_NULL;
  string pathAddHost;
  if (hosts.size() != 0) {
    if (rank == 0) {
      pathAddHost = "add-hosts-tmp";
      ofstream outfile(pathAddHost);
      for (auto host: hosts) {
        outfile << hosts[h] << std::endl;
      }
      outfile.close();
      MPI_Info_create(&info);
      MPI_Info_set(info, 
                   "add-hostfile", 
                   &pathAddHost[0]);
     }
  }
  MPI_Barrier(oldComm);
    
  int parentRank = 0;
  int changeRank = false;
  MPI_Comm interCommChildren;
  MPI_Comm_spawn(&childProgram[0],
                 MPI_ARGV_NULL,
                 numAdd,
                 MPI_INFO_NULL,
                 parentRank,
                 oldComm,
                 &interCommChildren,
                 NULL);
  MPI_Intercomm_merge(interCommChildren, 
                      changeRank, 
                      &newComm);
  MPI_Comm_free(&interCommChildren);
  if (info != MPI_INFO_NULL) {
    MPI_Info_free(&info);
    if (rank == 0) remove(&pathAddHost[0]);
  }
}
\end{lstlisting}
In Lines~4--20, host information from the API is passed to MPI via MPI\_Info.
In Lines~24--32, Rank \texttt{parantRank} generates new processes.
Since the generated communicator by the \texttt{MPI\_Comm\_spawn} is an inter-communicator, we need to convert it to an intra-communicator (Lines~33-35). 

\medskip

Code~\ref{lst:scaleoutnew} shows the implementation for the new process side. 
After getting the inter-communicator among the original processes, intra-communicator is created for direct communication among all the processes.
\begin{lstlisting}[caption={Scale Out Implementation in C++ (New Processes)},label={lst:scaleoutnew}]
void initNewProcess(MPI_Comm& newComm) {
  MPI_Comm interCommToParent;
  MPI_Comm_get_parent(&interCommToParent);
  int changeRank = true;
  MPI_Intercomm_merge(interCommToParent, 
                      changeRank, 
                      &newComm);
  MPI_Comm_free(&interCommToParent);
}
\end{lstlisting}

\newpage

\subsection{Scale In Implementation}
Code~\ref{lst:scalein} shows the implementation for scaling in. 
According to the flag \texttt{isRemoving}, Lines~7--27 manages the host information.
Then, in Lines~29--30 MPI processes are split into remaining processes and removal ones. The removal processes leave from the original communicator and terminate the program.
\begin{lstlisting}[caption={Scale In Implementation in C++},label={lst:scalein}]
void scaleIn(MPI_Comm& oldComm, bool isRemoving, 
    MPI_Comm& newComm, bool& thisCanTerminate) {
  int rank, size;
  MPI_Comm_rank(oldComm, &rank);
  MPI_Comm_size(oldComm, &size);
    
  char remainingHostname[CSIZE];
  if (!isRemove) {
    gethostname(remainingHostname, CSIZE);
  }
  string remainingHosts = string(CSIZE*size, 'N');
  MPI_Allgather(remainingHostname, 
                CSIZE, 
                MPI_CHAR, 
                &remainingHosts[0], 
                CSIZE, 
                MPI_CHAR, 
                oldComm);
  thisCanTerminate = true;
  char myhostname[CSIZE];
  gethostname(myhostname, CSIZE);
  for (int r = 0; r < size; ++r) {
    if (strncmp(&remainingHosts[r*CSIZE], 
        myhostname, CSIZE) == 0) {
      thisHostCanBeRemoved = false;
    }
  }

  MPI_Comm_split(oldComm, isRemoving,
                 rank, &newComm);
  if (isRemoving) {
    MPI_Comm_free(&oldComm);
    MPI_Comm_free(&newComm);
    MPI_Finalize();
  }
};
\end{lstlisting}

\section{Performance Evaluation}
In this section, we provide the empirical evaluation for our implementations.
We first summarize the set up for the evaluation. 
Then, we show the performance results for \texttt{ScaleOut()} and \texttt{ScaleIn()}.
\subsection{Set up}
We use four Ubuntu servers (ver 18.04). Each computing node is connected by 10 GbE and has dual sockets of Intel Xeon CPU E5-2697 v4 (18 cores per socket, 2.30GHz) with 500GB RAM.
The number of MPI slots in each node is set to be 32. Thus, we totally use 128 cores (32$\times$4) for MPI program. 
Open~MPI~2.1.1 is used and the program is compiled via GCC 9.1.0 with -O3 optimization.
All the experiments are conducted five times and the average values are illustrated.

\subsection{Results}
\begin{figure}[h]
\begin{minipage}[t]{.44\textwidth}
  \centering
  \includegraphics[width=\columnwidth]{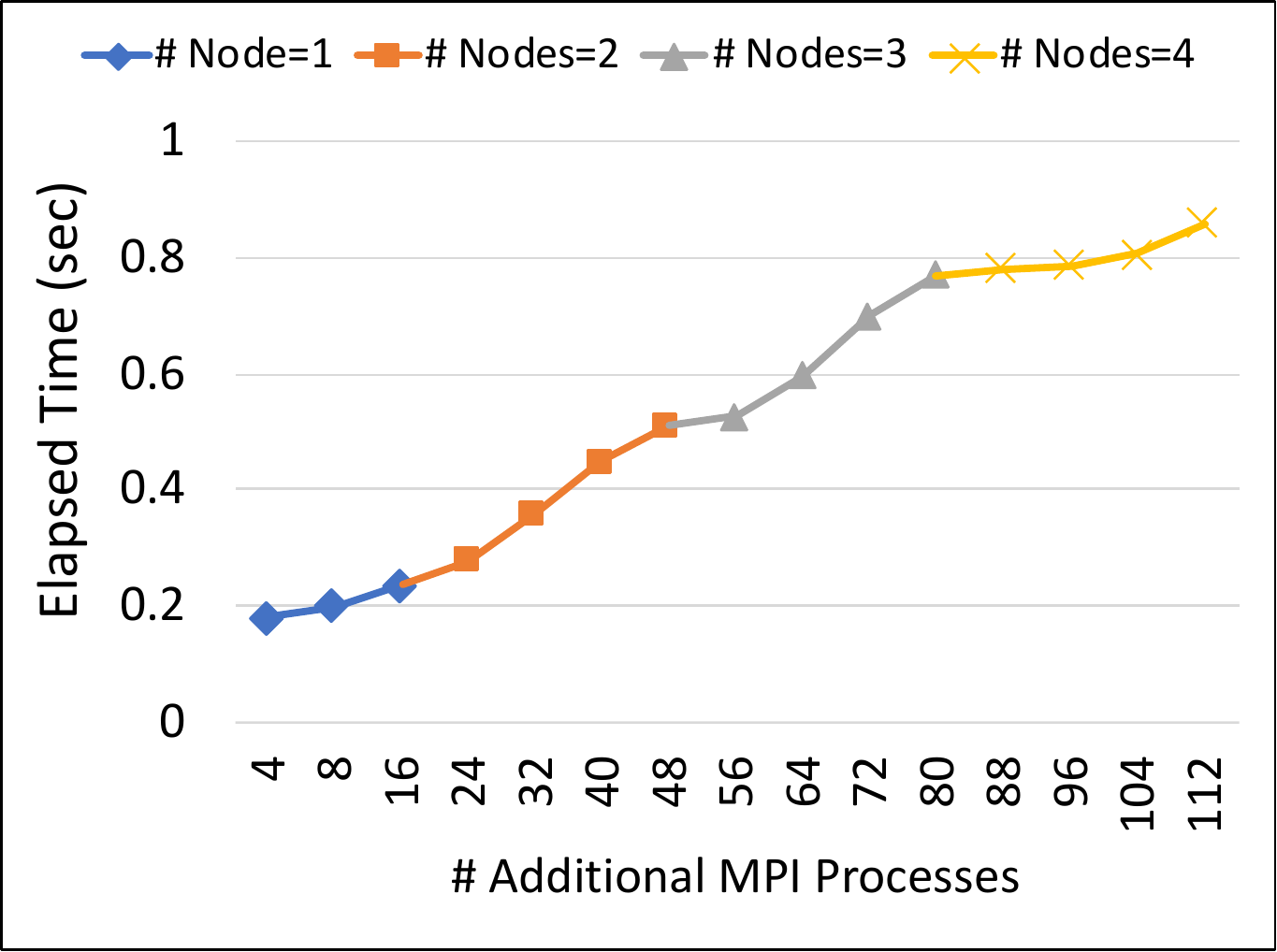}%
  \caption{Performance of \texttt{ScaleOut()}. \# Initial Processes is 16.}%
  \label{fig:per_add}
\end{minipage}
\hfill
\begin{minipage}[t]{.44\textwidth}
  \centering
  \includegraphics[width=\columnwidth]{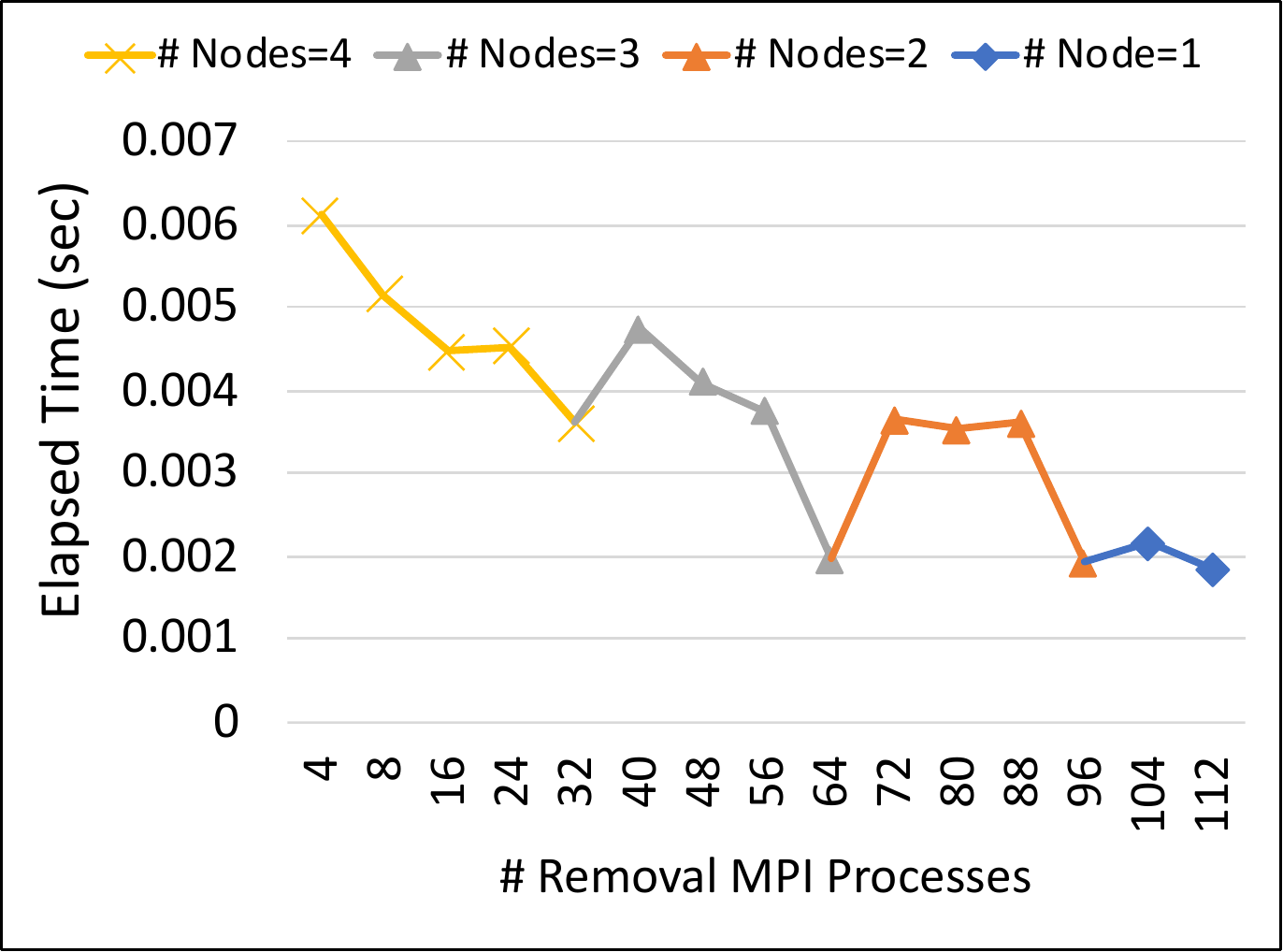}%
  \caption{Performance of \texttt{ScaleIn()}. \# Initial Processes is 128.}%
  \label{fig:per_rm}
\end{minipage}
\end{figure}

Figure~\ref{fig:per_add} shows the performance result of \texttt{ScaleOut()}.
In this scenario, 16 MPI processes are initially executed and 4 to 112 MPI processes are added during the execution.
Since each machine has 32 MPI slots, 4 to 16 additional processes are from the same computing node as the initial one; 24 to 48 processes are from the second node; 56 to 80 from the third node; 88 to 112 are from the last node.
Thus, for example in the case of 56 additional MPI processes in Figure~\ref{fig:per_add}, 3 computational nodes are totally used (the gray line).

Overall, \texttt{ScaleOut()} can be done within a second.
The elapsed time for \texttt{ScaleOut()} increases linearly as the number of the additional MPI processes.
Its gradient is slightly changed in each number of nodes. 
For example, from 3 to 4 nodes (i.e., from the gray to yellow line), its gradient becomes smaller.

\medskip

Figure~\ref{fig:per_rm} shows the performance result of \texttt{ScaleIn()}.
In this scenario, 128 MPI processes are initially executed before 4 to 112 MPI processes are removed during the execution.
Different line colors in Figure~\ref{fig:per_rm} represent the number of nodes for remaining processes after \texttt{ScaleIn()}
For example, in the case of 56 removal processes, 3 computing nodes are used for the remaining processes.

Overall, \texttt{ScaleIn()} is much faster than \texttt{ScaleOut()}.
Its elapsed time roughly decreases as the number of removal MPI processes. 
This is due to the fact that establishing the new intra-communicator among the remaining processes is dominant for the workload. Basically, its elapsed time increases as the number of remaining processes, except that it increases at the point to change the number of the computing nodes (for example from 32 to 40 in Figure~\ref{fig:per_rm}). 

Table~\ref{tb:perf} shows the performance breakdown of \texttt{ScaleOut()}.
We measure the performance of \texttt{MPI\_Comm\_spawn()} and the other operations.
In \texttt{ScaleOut()}, the most part of the elapsed time is caused by \texttt{MPI\_Comm\_spawn()}.
The establishment of the intra-communicator and the management of host information are much faster and only a few parts of \texttt{ScaleOut()}. 

\begin{table}[H]
\centering
\caption{Performance Breakdown of \texttt{ScaleOut()} (sec).}\label{tb:perf}
\begin{tabular}{lrrrr} \hlinewd{2\arrayrulewidth}
                               & \# Add = 16  & 48       & 80     & 112    \\ \hline
\texttt{ScaleOut()}            &  0.2375      & 0.5110   & 0.7705 & 0.8611 \\
-- \texttt{MPI\_Comm\_spawn()} &  0.2226      & 0.4114   & 0.6681 & 0.7582 \\ 
-- The Other Operations        &  0.0149      & 0.0996   & 0.1024 & 0.1029 \\ \hlinewd{2\arrayrulewidth}
\end{tabular}
\end{table}

\section{Conclusion}
In this paper, we present the implementation tips for dynamic scaling on MPI. 
Our implementation does not only create new processes but also establish an intra-communicator among all the processes and manage the host information.
We report the performance trends for the dynamic scaling, showing that the overhead of the dynamic scaling is acceptable on 16 to 128 MPI processes although it is not too small to be completely ignorable.
Our implementation is published as the library to be able to easily extend an MPI program into a dynamic-scalable one.

\bibliographystyle{unsrt}

\bibliography{ref}

\end{document}

%% file: packages.tex
\usepackage{booktabs} 
\usepackage{graphicx}
\usepackage{here}
\usepackage[ruled, linesnumbered, vlined]{algorithm2e}
\usepackage{subfigure}
\usepackage{calc}
\usepackage{tabularx}
\usepackage{hyperref}
\usepackage{balance}
\usepackage{mathtools}
\usepackage{lipsum,multicol}
\usepackage{soul}

\usepackage{amsmath}

\mathchardef\mhyphen="2D 

\usepackage{amsthm}
\newtheorem*{theorem*}{Theorem}

\theoremstyle{acmplain}

\theoremstyle{acmplain}

\usepackage{xcolor}

\usepackage{array,multirow}
\usepackage{float}

\usepackage{listings}

\makeatletter
\def\hlinewd#1{%
\noalign{\ifnum0=`}\fi\hrule \@height #1 %
\futurelet\reserved@a\@xhline}
\makeatother

\makeatletter
\newcommand{\removelatexerror}{\let\@latex@error\@gobble}
\makeatother
\let\oldnl\nl
\newcommand{\nonl}{\renewcommand{\nl}{\let\nl\oldnl}}%

\usepackage[top=4cm, bottom=4cm, left=4cm, right=4cm]{geometry}

%% file: main.bbl
\begin{thebibliography}{1}

\bibitem{spotinstance}
{Amazon EC2 Spot Instances}.
\newblock \url{https://aws.amazon.com/ec2/spot/}.

\bibitem{preemptible}
{Google Preemptible VMs}.
\newblock \url{https://cloud.google.com/preemptible-vms/}.

\bibitem{mpi2}
{MPI-2.0 documents}.
\newblock
  \url{https://www.mpi-forum.org/docs/mpi-2.0/mpi-20-html/mpi2-report.html}.

\end{thebibliography}
